\documentclass[aps,pre,twocolumn,amsmath,amssymb,nofootinbib,superscriptaddress,floatfix,eqsecnum]{revtex4-1}

\usepackage{amsmath,amsfonts,amssymb,eufrak}
\usepackage{amssymb}
\usepackage{amsthm}
\usepackage[dvipdf]{color}
\usepackage{graphicx}
\usepackage{dcolumn}
\usepackage{bm}

\begin{document}
\title{Topological mechanics in quasicrystals}

  \author{Di Zhou}
 \affiliation{
 Department of Physics,
  University of Michigan, Ann Arbor, 
 MI 48109-1040, USA
 }

  \author{Leyou Zhang}
 \affiliation{
 Department of Physics,
  University of Michigan, Ann Arbor, 
 MI 48109-1040, USA
 }

  \author{Xiaoming Mao}
 \affiliation{
 Department of Physics,
  University of Michigan, Ann Arbor, 
 MI 48109-1040, USA
 }

\begin{abstract}
We study topological mechanics in two-dimensional quasicrystalline parallelogram tilings.  Topological mechanics has been studied intensively in periodic lattices in the past a few years, leading to the discovery of topologically protected boundary floppy modes in Maxwell lattices.  In this paper we extend this concept to quasicrystalline parallelogram tillings and we use the Penrose tiling as our example to demonstrate how these topological boundary floppy modes arise with a small geometric perturbation to the tiling.  The same construction can also be applied to disordered parallelogram tilings to generate topological boundary floppy modes.  
  We prove the existence of these topological boundary floppy modes using a duality theorem which relates floppy modes and states of self stress in parallelogram tilings and fiber networks, which are Maxwell reciprocal diagrams to one another.  We find that, due to the unusual rotational symmetry of quasicrystals, the resulting topological polarization can exhibit orientations not allowed in periodic lattices.  Our result reveals new physics about the interplay between topological states and quasicrystalline order, and leads to novel designs of quasicrystalline topological mechanical metamaterials.
\end{abstract}

\maketitle
\section{Introduction}

The notion of ``topological protection'' has far-reaching influences in modern condensed matter physics, from protected conducting states on the surface of topological insulators, to stable pinning of a superconductor above a magnet due to vortices which are topological defects in the complex order-parameter field.  Topology is concerned with properties of a system that are invariant under continuous deformations.  If a physical property is determined by topology, it is highly robust and protected against disorder, as long as the disorder is not strong enough to destroy the entire topological state.  
It is thus interesting to explore topologically protected phenomena in systems that are not periodic in space, which not only demonstrate the power of topological protection under strong disorder, but may also offer new physics that was not present in periodic lattices.

In this paper, we investigate topological mechanics in two-dimensional (2D) quasicrystalline structures.  Topological mechanics is a very active new research direction applying ideas of topological states of matter to classical mechanical networks~\cite{Prodan2009,Kane2014,Lubensky2015,Wang2015,Nash2015,Suesstrunk2015,Mousavi2015,
Yang2015,Peano2015,Strohm2005,Sheng2006,Pal2016,He2016,Suesstrunk2016,
Xiao2015a,Rocklin2016,Rocklin2017,Paulose2015a,Paulose2015,Chen2014,mao2018maxwell}.  We focus on a peculiar branch of topological mechanics concerning ``Maxwell lattices'' which are mechanical lattices with balanced numbers of degrees of freedom and constraints [$\langle z \rangle =2d$ where $\langle z \rangle$ is the average coordination number and $d$ is the spatial dimension] and are thus on the verge of mechanical instability~\cite{Kane2014,Lubensky2015,Rocklin2016,Rocklin2017,Paulose2015a,
Paulose2015,Chen2014,Rocklin2014,Feng2016,Zhang2016,mao2018maxwell,Zhou2018,Zhang2018}.  Studies of topological mechanics in Maxwell lattices has led to topologically protected surface modes at zero frequency, which dictate local stiffness of the system.  So far most investigations of topological mechanics are done in periodic lattices, with a few exceptions on systems such as jammed packings~\cite{Sussman2016}, amorphous networks of gyroscopes~\cite{mitchell2018amorphous}, active fluids~\cite{souslov2018topological}, 
disordered fiber networks~\cite{Zhou2018}, patterns~\cite{apigo2018topological}, etc.

Quasicrystals are fascinating materials characterized by long-range orientational order and quasiperiodic (rather than periodic) translational order~\cite{penrose1974role,levine1984quasicrystals,elser1985crystal,
levine1986quasicrystals,socolar1986quasicrystals,steinhardt1987physics}.  They fall between ordered periodic lattices and disordered structures, and offer us a great arena where we can study topological protection in systems lack of translational symmetry but allow analytic treatment.  Furthermore, quasicrystals exhibit intriguing physics that was not available in crystals, such as rotational symmetries that are forbidden in periodic lattices, physics of higher dimensions, and self-similarity.  The interplay of these unique features with topological states of matter can offer a rich variety of interesting phenomena.  Some of these phenomena have been discussed in the context of photonic quasicrystals~\cite{Kraus2012,bandres2016topological}.  

\begin{figure*}[t]
\centering
\includegraphics[width=1\textwidth]{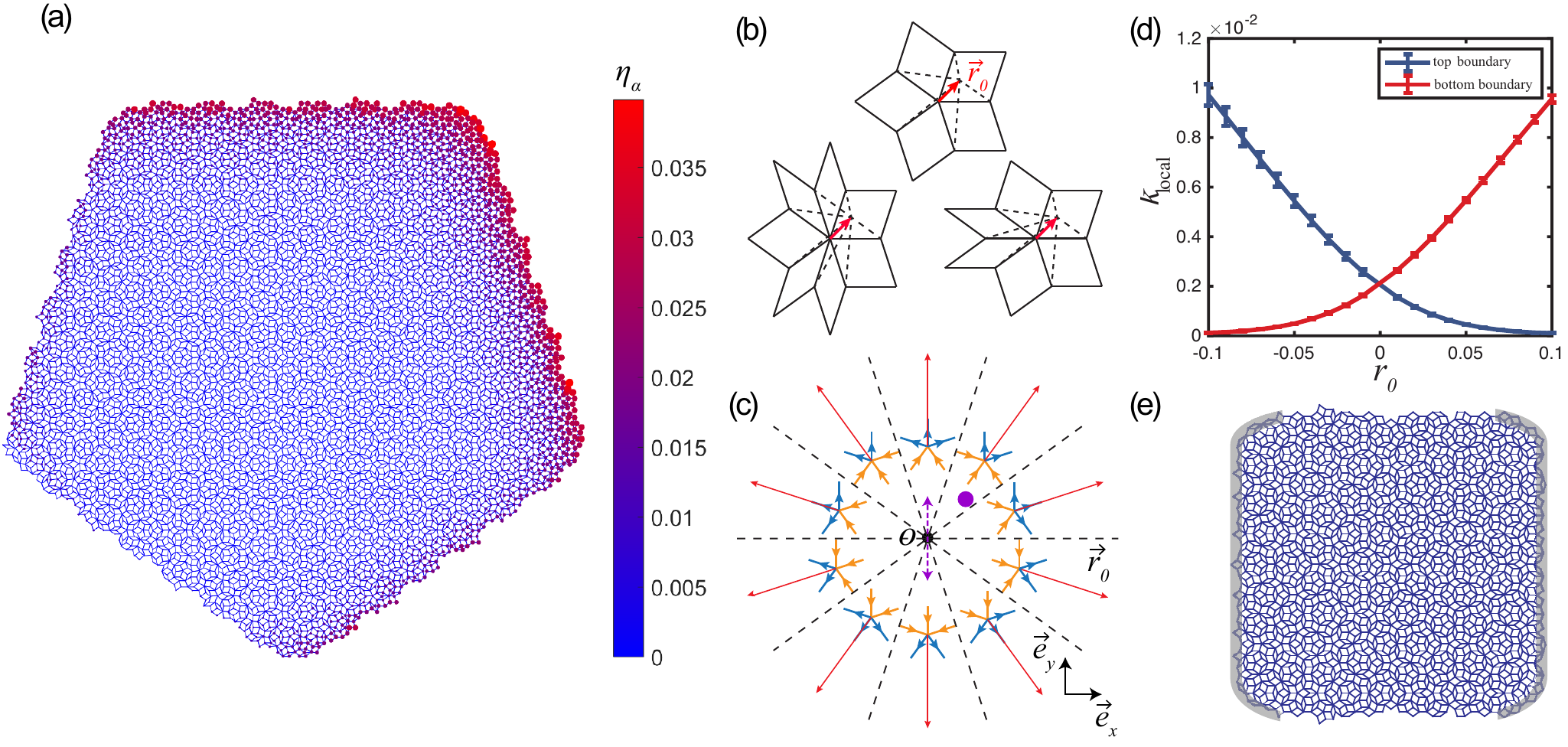}
\caption{Topological mechanics of the Penrose tiling.  (a) A topologically polarized Penrose tiling with $\vec{R}_T$ pointing towards the upper-right tip of the pentagon ($54^\circ$ from $x$ axis).  Zero modes weight $\eta_{\alpha}$ (defined in Sec.~\ref{SEC:TopoQC}) on each site $\alpha$ is shown using both color (see color bar on the right) and size of the disks.  The bottom left boundary, the outward normal of which is opposite to $\vec{R}_T$, is free of floppy modes. (b) Geometric changes we apply on the Penrose tiling to induce the topological boundary floppy modes shown in (a).  We choose these $z=5,6,7$ vertices (and ones related to these by rotation) in the Penrose tiling and displace them by a small amount $\vec{r}_0 = r_0 \hat{e}_{\textrm{disp}}$ with $r_0=0.15\times$length of edges in the Penrose tiling, 
and $\hat{e}_{\textrm{disp}} = (\cos \psi, \sin \psi)$ with $\psi=45^\circ$.  Red arrows show an example of such displacements ($r_0$ magnified to be visible), which leads to (a).  
(c) Phase diagram of the resulting topological polarization of strips of parallelograms in the five symmetry  directions (short blue out arrow for $n_i=1$ and short orange in arrows for $n_i=-1$ where $n_i$ is a topological invariant defined in Sec.~\ref{SEC:TopoQC} for strip topological polarizations), and the topological polarization of the whole modified Penrose tiling $\vec{R}_T$ (long red arrows) as given by Eq.~\eqref{EQ:RT}, as these vertices are displaced.  When $\hat{e}_{\textrm{disp}}$ is in each of the 10 ranges of directions defined by the dashed lines, the polarizations are shown by the arrows in that range.  The purple dot shows the direction for the displacements used in (a) which has $\psi=45^\circ$, leading to $\vec{R}_T$ pointing towards $54^\circ$.  (d) Measured local stiffness $k_{\textrm{local}}$ (defined in Sec.~\ref{SEC:TopoQC} and App.~\ref{APP:Simulation}) on opposite boundaries of a topologically polarized Penrose tiling.  The setup of the simulation is shown in (e) where we cut a square piece of a polarized Penrose tiling and hold the left and right boundaries  fixed (shaded regions).  The tiling is polarized by displacing the $z=5,6,7$ vertices shown in (b) with $\psi=90^\circ$ (purple arrows in (b) corresponding to $r_0>0$ and $r_0<0$), leading to $\vec{R}_T$ pointing up for $r_0>0$ and down for $r_0<0$.  The measured local stiffness (d) shows that the boundary $\vec{R}_T$ points toward is dramatically softer than the opposite boundary.
}
\label{FIG:TopoQCExample}
\end{figure*}

Our paper focuses on topological mechanics in quasicrystals and we find that topological boundary floppy modes can be generated in 2D quasicrystalline structures by infinitesimal changes in the geometry.  We use the Penrose tiling, a well-known quasicrystalline structure composed of two types of parallelograms, as our example to demonstrate these topological boundary floppy modes.  In fact, our results apply to all parallelogram-tilings in 2D, including periodic, quasiperiodic, and disordered ones.  As we show below, all parallelogram-tilings have mean coordination $\langle z \rangle =2d$ and are thus Maxwell networks (note that we limit ourselves to tilings where all edges are ``complete'', i.e., nodes of a parallelogram merge with nodes of the neighboring parallelogram when they are tiled together, instead of sitting in the middle of the edge of other parallelograms).  Here we extend the notion of ``Maxwell lattices'' to``Maxwell networks'' to include aperiodic networks with $\langle z \rangle =2d$~\cite{Souslov2009,Mao2010,Ellenbroek2011,Mao2011a,Sun2012,Zhang2015a,Mao2015}.  Jammed packings of frictionless spheres~\cite{Liu2010} and Mikado fiber networks~\cite{Head2003,Wilhelm2003} are both examples of Maxwell networks.  We show that original parallelogram-tilings have bulk floppy modes.  Small changes in the geometry can topologically polarize these parallelogram tilings and transform these bulk floppy modes into boundary floppy modes, as shown in Fig.~\ref{FIG:TopoQCExample} (a-b).

The way we prove the existence of these boundary floppy modes is via a duality theorem relating floppy modes and states of self stress in a mechanical network and its Maxwell reciprocal diagram~\cite{crapo1994spaces,mitchell2016mechanisms}.  We find that, parallelogram tilings and fiber networks are Maxwell reciprocals of one another, and based on previous studies of topological boundary floppy modes in fiber networks, we construct methods to topologically polarize parallelogram tilings.

A particularly interesting property of topological quasicrystalline structures is that they can display rotational symmetries that are not allowed in periodic structures.  As a result, the topological polarization $\vec{R}_T$ of a given quasicrystalline structure can be tuned to point to more directions than in periodic lattices [Fig.~\ref{FIG:TopoQCExample} (c)], as we discuss in details below.  This can be viewed as a reflection of topological states in higher dimensions, and brings interesting new physics to topological mechanics.  

A consequence of these topological boundary floppy modes is that topologically polarized Penrose tilings show contrasting local stiffness at its boundaries where  $\vec{R}_T$ points toward or away from [Fig.~\ref{FIG:TopoQCExample} (d-e)].  This can lead to interesting new designs of mechanical metamaterials with quasicrystalline structures and unusual rotational symmetries of topological polarizations.

\section{Quasicrystalline structures and their floppy modes}
\subsection{Methods of generating quasicrystalline structures}
In this section we briefly review two well known methods to generate quasicrystalline structures, the cut-and-project method (CPM)~\cite{de1981algebraic} and the generalized dual method (GDM)~\cite{socolar1985quasicrystals}, both of which are very useful for our later discussions of topological mechanics in quasicrystals.
 
In the CPM, a quasicrystalline structure is obtained by cutting a periodic lattice in a higher dimensional ($D$) space with a lower dimensional ($d$) surface that is incommensurate with the lattice planes, and project lattice sites that are within a certain distance to the surface.  

The Penrose tiling can be obtained by cutting a five-dimensional (5D) hypercubic lattice with a 2D plane.  The 2D plane is spanned by two vectors in the 5D space $\vec{e}_1$ and $\vec{e}_2$, with their components in the 5D space given by $e_{1i} = \cos (i-1)\theta_p$, $e_{2i} = \sin(i-1)\theta_p$, for $i=1,2,\ldots,5$, and $\theta_{\textrm{p}} = 2\pi/5$ ($\textrm{p}$ stands for Penrose tiling).  Each site in the 5D hypercubic lattice can be labeled by a set of five integers $\vec m = (m_1,m_2,\ldots,m_5)$ and we take the lattice constant to be 1.  Now we pick out a thin layer of lattice sites very close to the 2D plane spanned by $\vec{e}_1$ and $\vec{e}_2$ according to the following equation,
\begin{eqnarray}\label{EQ:mFloor}
m_i = \left \lfloor s_1 e_{1i}+s_2 e_{2i}+\gamma_i+\frac{1}{2} \right \rfloor
\end{eqnarray}
where $s_1,s_2 \in \mathbb{R}$, making the combination $s_1 e_{1i}+s_2 e_{2i}$ running through all points in the 2D plane.  $\left \lfloor x\right \rfloor$ is the integer floor function which gives the integer part of $x$,  $\gamma_i$'s are offset constants, and the term $1/2$ shifts the floor function such that it gives the closest integer.

\begin{figure}[h]
\centering
\includegraphics[width=0.3\textwidth]{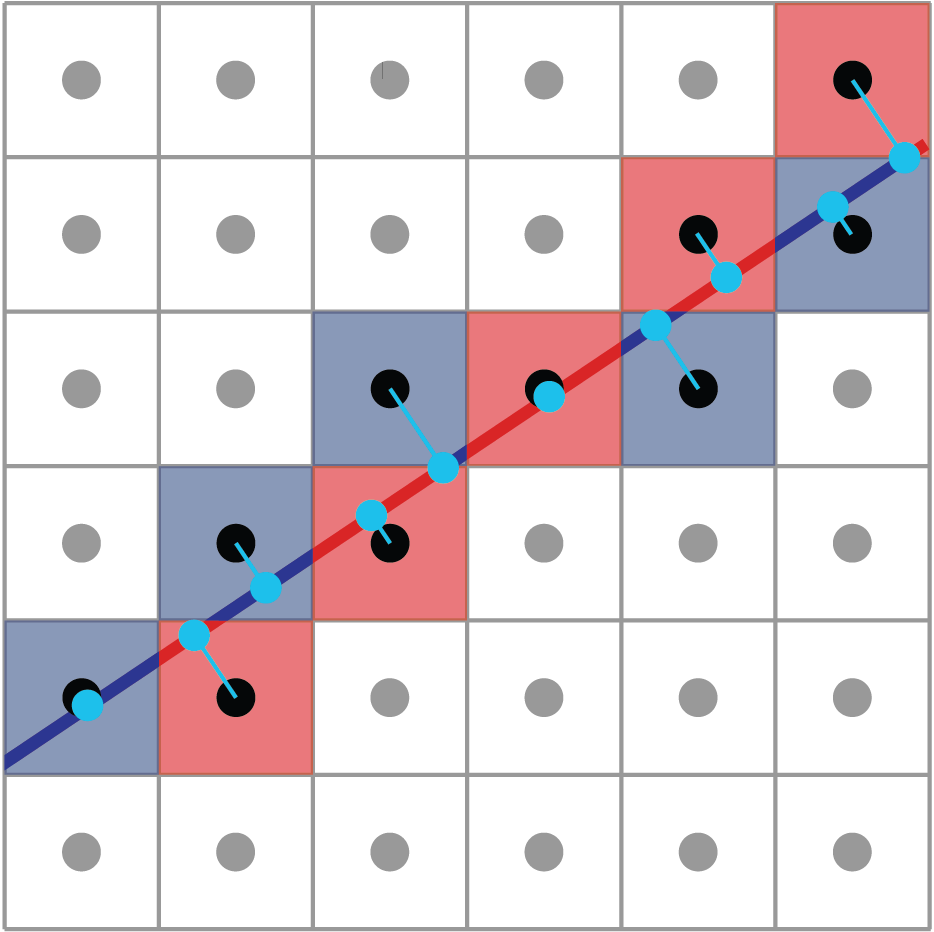}
\caption{Using the CPM to project a 2D square lattice to a 1D quasicrystalline chain.  The solid line cuts the square lattice in an incommensurate direction.  It is separated into small segments which belong to unit cells it passes through (represented by red and blue colors).  Each segment is mapped into the lattice site (black disks) associated with the unit cell it belongs to [according to the 1D version of Eq.~\eqref{EQ:mFloor}].  These sites are then projected to the line, yielding the quasicrystalline 1D chain (cyan disks).
}
\label{FIG:CPM}
\end{figure}

This equation maps each point on the plane into a thin layer of sites in the 5D lattice.  This is a non-injective mapping from a continuous plane onto discrete sites.  An intuitive way to think about this mapping is to assign each site in the 5D lattice a hypercube unit cell centered around this site.  The part of the cutting plane in a unit cell is mapped into the site of this unit cell.  A 2D to 1D mapping of this type is shown in Fig.~\ref{FIG:CPM}.  Taking other values of $\gamma_i$'s simply shifts the cutting plane.  

Projecting these chosen sites (each labeled by an array of five integers $\vec{m}$) onto the 2D plane yields a set of points
\begin{eqnarray}\label{EQ:rCPM}
\vec r_{\vec{m}} =\{ \vec m\cdot \vec{e}_1 , \vec m \cdot \vec{e}_2 \} ,
\end{eqnarray}
which are sites in the Penrose tiling.  Edges in the Penrose tiling are defined between pairs of projected sites that were nearest neighbors in the 5D lattice.

\begin{figure}[h]
\centering
\includegraphics[width=0.35\textwidth]{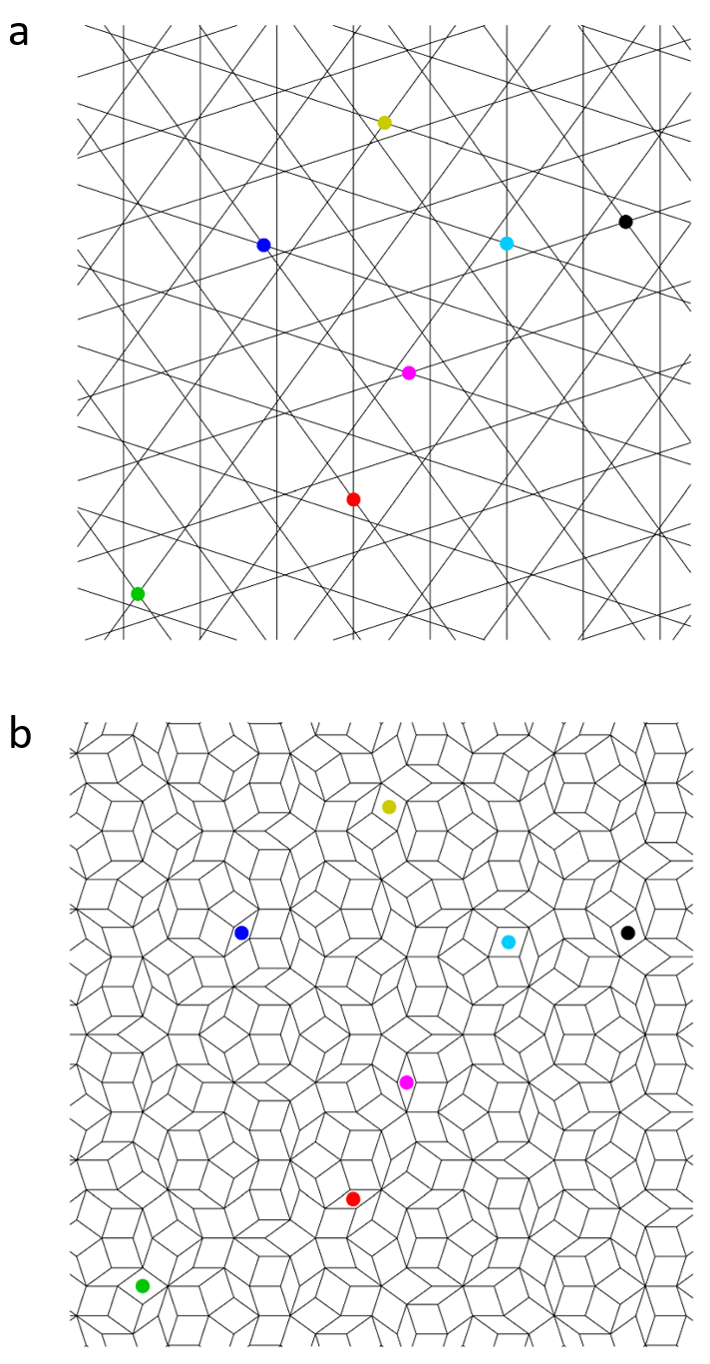}
\caption{Duality between a $5$-grid and the Penrose tiling it generates through the GDM.  A few nodes in the $5$-grid and their corresponding parallelograms in the Penrose tiling are marked by the same color.
}
\label{FIG:GDMPenrose}
\end{figure}

The GDM is a more general method that can be used to generate quasicrystals with arbitrary orientational symmetry.  This method consists of the following five steps: (1) $D$ ``star-vectors'', $\vec{a}_i$ with $i=1,\ldots,D$, are chosen in a $d$-dimensional space ($d=2$ in the cases we consider).  (2) An infinite set of periodically or quasiperiodically spaced parallel planes are introduced normal to each star vector, forming a $D$-grid.  (3) Each plane normal to $\vec{a}_i$ is labeled by an integer $m_i$ representing its ordinal position in the $\vec{a}_i$ direction.  (4) The $d$-dimensional space is cut by the planes into non-overlapping polyhedra which can be labeled by a set of $N$ integers $(m_1,\ldots,m_D)$ such that the polyhedra is between planes $m_i-1$ and $m_i$ in each direction $\vec{a}_i$. (5) Each polyhegra is then mapped into a site in the quasicrystal through
\begin{align}\label{EQ:rGDM}
	\vec r_{\vec{m}} = \sum_{i=1}^{D} m_i \vec{a}_i .
\end{align}
The special example of Penrose tiling is obtained by taking $D=5$ and star vectors in directions of $0, 2\pi/5, 4\pi/5, 6\pi/5, 8\pi/5$ in a plane ($d=2$) and the parallel planes are simply parallel lines, which cut the 2D plane into polygons, as shown in Fig.~\ref{FIG:GDMPenrose}.

The GDM reveals an intriguing dual relation between the Penrose tiling and $D$-grids which are networks of fibers.  Viewing these two structures as graphs, it is straightforward to see that there is a dual relation between them where edges $\leftrightarrow$ edges, and sites $\leftrightarrow$ faces (Fig.~\ref{FIG:GDMPenrose}).  Moreover, an edge in the Penrose tiling is perpendicular to the corresponding edge in the $D$-grid fiber network that is used to generate it, because edges in the CPM are along the star-vectors.  As we discuss in later sections, this dual relation makes the Penrose tiling and the $D$-grid fiber network generating the Penrose tiling Maxwell reciprocal diagrams of each other, and is the basis for establishing topological mechanics in the Penrose tiling, and more generally, any parallelogram tilings, from known results of the topological mechanics of fiber networks.

Quasicrystals generated by the CPM can all be generated by the GDM.  Their equivalence can be realized by viewing the parallel planes in the GDM as the projections of the $D$ families of the hypercubic lattice planes that separate the unit cells in the high dimensional crystal.  The polygons in the GDM correspond to the part of the cutting plane that belong to different unit cells of the lattice.  Equation~\eqref{EQ:rGDM} is equivalent to Eq.~\eqref{EQ:rCPM} because $\vec{a}_i$'s  are simply the projections of the $D$-dimensional lattice primitive vectors onto the $d$-dimensional space.

In practice, we generate our quasicrystalline parallelogram tilings using the GDM, which can be conveniently written into a set of simple equations to solve for $\vec{m}$.  Details of our numerical methods can be found in App.~\ref{APP:GDM}.

\subsection{Floppy modes in parallelogram-tilings}\label{SEC:BulkModes}
It is straightforward to construct floppy modes in tilings of parallelograms, given to their special geometry.  Floppy modes (FMs) are defined as normal modes of deformation of a structure that cost no elastic energy.  In this paper, we consider elastic networks of point-like particles (nodes as free hinges) connected by central-force springs (edges), so FMs are normal modes which do not change the length of any edge.  

\begin{figure}[h]
\centering
\includegraphics[width=0.4\textwidth]{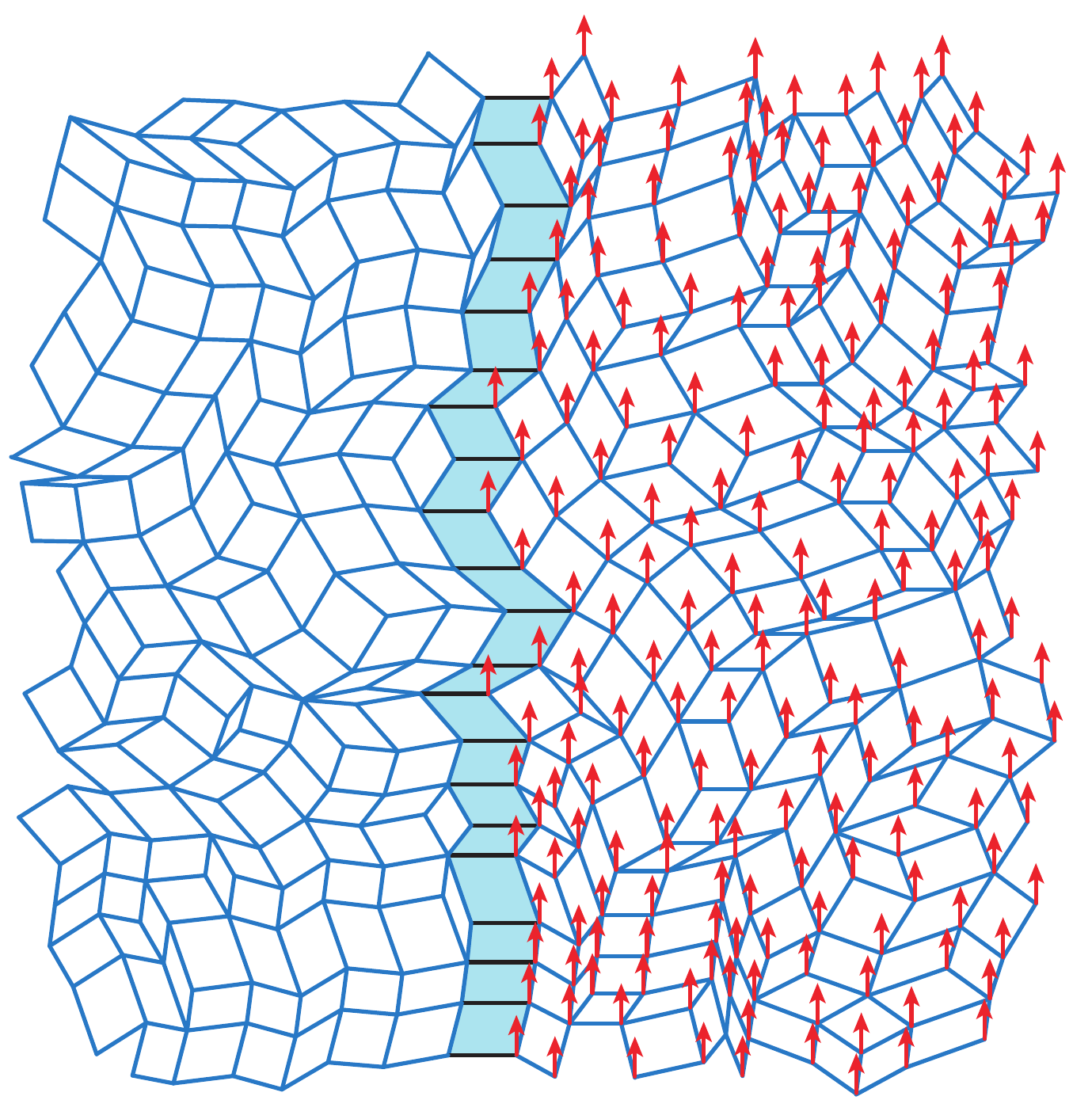}
\caption{An example of a bulk FM (red arrows) of a random parallelogram tiling. A strip of parallelograms (light blue) is randomly chosen and the FM displaces all nodes right to this strip in a direction that is perpendicular to the parallel edges in this strip.}\label{FIG:BulkMode}
\end{figure}

In a tiling of parallelograms, one can always start from an arbitrary parallelogram and uniquely define two ``strips'' of parallelograms following the two directions of parallel edges in this parallelogram.   Figure~\ref{FIG:BulkMode} shows an example of one such strip.  Parallelograms in each strip share parallel edges to one another.  
In the case of the Penrose tiling, such a strip simply corresponds to a set of parallelograms that come from all crosslinking points along the same fiber in the GDM.

Each strip separates the whole tiling into two parts, and a bulk FM immediately follows from a strip: one can hold the part left to this strip fixed, and pull the right side up by deforming parallelograms in the strip.  The displacement vectors, which are the same for all sites right to the strip, are chosen to be perpendicular to the parallel edges in the strip, so no edge lengths are changed.  This is a bulk mode because the magnitude of this mode does not decay from edge to the bulk.  
This FM actually extends to nonlinear order and leads to a finite mechanism of the structure.

The number of such bulk FMs is subextensive to the area of the tiling.  This follows from a simple counting of degrees of freedom and constraints.  For an infinite tiling of parallelograms (with complete edges), one has $N=F$ where $N,F$ are the numbers of nodes and faces, one can also write down Euler's formula $N+F-E=2$ (for open boundary where the exterior face is included) or $N+F-E=0$ (for periodic boundary conditions).  Thus, $\langle z \rangle = 2E/N = 4 $ which is exact for periodic boundary conditions and ignoring an additive constant of $O(1/N)$ for open boundary.  This tells us that parallelogram-tilings are Maxwell networks, meaning that they have balancing degrees of freedom and constraints in the bulk.  For a finite piece of parallelogram-tiling with open boundaries, the cut edges on the boundary give rise to FMs, the number of which is proportional to the size of the boundary.  More precisely, the total number of FMs is equal to the number of strips in a parallelogram-tilings minus one (given a properly cut boundary), as we discuss in more details in Sec.~\ref{SEC:TopoQC}.

In this sense, parallelogram-tilings are very similar to the classical Mikado fiber networks, where all FMs are bulk FMs as well.  In Ref.~\cite{Zhou2018} we showed that with a small change in the geometry, Mikado fiber networks can be topologically polarized, where bulk FMs becomes topological boundary FMs.  In what follows, we show that a similar geometric change can be done in the parallelogram-tilings as well, leading to topological boundary FMs.

\section{Mechanical duality between parallelogram-tilings and fiber networks}\label{SEC:DualMech}
In this section, we review the concept of Maxwell reciprocal diagrams (which we abbreviate as ``reciprocal diagrams'' below) and show that parallelogram-tilings and fiber networks are reciprocal diagrams of each other.  We further review the mechanical duality between reciprocal diagrams, which relate their FMs and states of self stress respectively.

\subsection{Maxwell reciprocal diagrams and equilibrium stresses}\label{SEC:MaxwellReciprocal}
J.~C.~Maxwell introduced the concept of reciprocal diagrams (called ``reciprocal figures'' in his original papers) and used it to solve equilibrium forces on mechanical frames~\cite{maxwell1864xlv,maxwell1870reciprocal}.  Two diagrams $A$ and $A^*$, which are both networks of nodes connected by straight edges, are reciprocal to one another if:
\begin{itemize}
\item they contain equal numbers of edges;
\item corresponding edges in the two diagrams are perpendicular to one another;
\item corresponding edges that converge to a point in one diagram form a closed polygon in the other.
\end{itemize}
Per Maxwell~\cite{maxwell1864xlv,maxwell1870reciprocal}, ``reciprocal figures are such that the properties of the first relative to the second are the same as those of the second relative to the first.''  Here the two reciprocal diagrams have a dual relation between them where edges $\leftrightarrow$ edges, and sites $\leftrightarrow$ faces, similar to the relation between dual graphs, but with the extra requirement that corresponding edges are perpendicular.  In some versions, reciprocal diagrams are also defined with parallel edges, but they simply relate to reciprocal diagrams with perpendicular edges by a homogeneous rotation of $\pi/2$.

\begin{figure*}[t]
\centering
\includegraphics[width=1\textwidth]{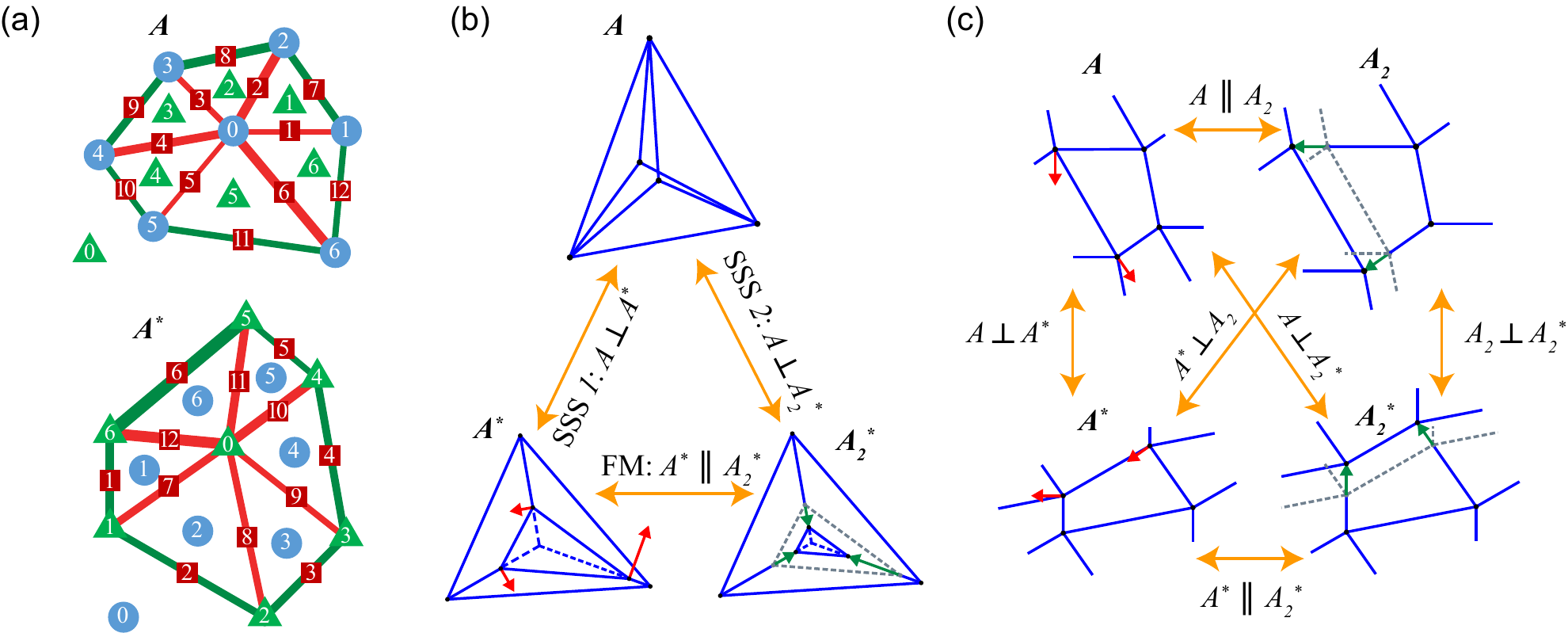}
\caption{Maxwell reciprocal diagrams and the mechanical duality theorem.  (a) A pair of reciprocal diagrams $A\perp A^*$.  Edges (marked by squares) corresponding to one another are labeled by the same number in $A$ and $A^*$.  Nodes in $A$ (circles) correspond to faces in $A^*$ (including the exterior face) with the same number.  Faces in $A$ (triangles) correspond to nodes in $A^*$ with the same number.  
Equilibrium stresses that generate the reciprocal is marked by the thickness of the edges and the red and green colors correspond to two signs of tension.
(b) Two SSSs of frame $A$ leads to its two reciprocal diagrams $A^*, A^*_2$.  This shows an example of the mechanical duality theorem discussed in Sec.~\ref{SEC:DualityTheorem}: FMs of $A^*$ and SSSs of $A$ (except for the one that generates $A^*$) have a one-to-one mapping---they both correspond to $A^*_2$.  The irrotational and FM displacements $\vec{v}$ and $\vec{u}$ are shown by green and red arrows respectively (defined in Sec.~\ref{SEC:DualityTheorem}).  
(c) Reciprocal relations between periodic lattices.  Any Maxwell lattice $A$ must have at least two SSSs at $\vec{q}=0$ [as discussed at Eq.~\eqref{EQ:NsLatt}], leading to two reciprocal diagrams $A^*$ and $A_2^*$, which are both Maxwell lattices as well with the same periodicity.  Each of  $A^*$ and $A_2^*$ must also have at least two SSSs, leading to $A$ and $A_2$.  As discussed in Sec.~\ref{SEC:DualityTheorem}, because $A\parallel A_2$ and $A^* \parallel A_2^*$, they are related by irrotational displacements $\vec{v}$ (green arrows) and the corresponding FM $\vec{u}$ (red arrows) which belong to a group of FMs in Maxwell lattices called the Guest-Hutchinson FMs (see Ref.~\cite{mao2018maxwell}).
}\label{FIG:reciprocalOne}
\end{figure*}

As pointed out by Maxwell~\cite{maxwell1864xlv,maxwell1870reciprocal}, starting from a frame $A$, a reciprocal diagram $A^*$ can be built from an equilibrium tension distribution on edges (struts) in $A$.  Here ``equilibrium'' refers to the condition that the net force on any node (hinge) is zero.  The length of each edge $i^*$ in the reciprocal diagram $A^*$ is proportional to the tension $t_i$ on the corresponding edge $i$ in the original frame $A$.  The condition that the total force is zero on each node in $A$ can be written as
\begin{align}
	\sum_{i}^{\textrm{node } \alpha} t_{i} \hat{b}_{i} =0
\end{align}
where the sum is over all edges $i$ that connect to node $\alpha$ and $\hat{b}_{i} $ is the unit vector along the edge $i$.  Note that the tension $t_i$ can have positive or negative signs, and the edge length will be just determined by the magnitude $\vert t_i \vert$.  A convention can be taken such that one first assigns a direction $\hat{b}_{i}$ to every edge and if the force on node $\alpha$ is positive if it is along $\hat{b}_{i}$, and negative if it is against $\hat{b}_{i}$.
As a result the corresponding set of edges $i^*$ (of length $\vert t_i\vert$ and direction $\textrm{sgn}(t_i)\hat{b}_i^* \perp \hat{b}_i$) form a closed polygon $\alpha^*$ in the reciprocal diagram $A^*$, yielding a face that is dual to node $\alpha$ in $A$.  Each edge $i$ in $A$ connects two nodes $\alpha, \beta$, which correspond to two faces $\alpha^*, \beta^*$ that share edge $i^*$ in $A^*$.  In addition, 
 one can also view $A^*$ as a mechanical frame and $A$ as the reciprocal diagram, due to their reciprocity.  From this geometric relation, it is obvious to see that the converse of this statement is also true: if a frame has a reciprocal diagram, it must be able to carry an equilibrium distribution of stress.  In Fig.~\ref{FIG:reciprocalOne}(a) we show an example of a pair of reciprocal diagrams.  In the following discussion we will denote such reciprocal relation as $A\perp A^*$.

After the concept of ``state of self stress'' was introduced~\cite{Calladine1978,PellegrinoCal1986}, a one-to-one correspondence between reciprocal diagrams and states of self stress has been established rigorously~\cite{crapo1994spaces,mitchell2016mechanisms}.  
Therefore, the existence of reciprocal diagrams for a given frame is equivalent to the existence of states of self stress.

In what follows we review these mechanical concepts and discuss a mechanical duality theorem that we use to introduce topological boundary FMs in parallelogram tilings.

\subsection{States of self stress, floppy modes, and the existence of dual diagrams}\label{SEC:CountDual}
Linear mechanical properties of frames can be described by the equilibrium matrix $\mathbf{Q}$ which controls the statics and the compatibility matrix $\mathbf{C}$ which controls the kinetics~\cite{Calladine1978,PellegrinoCal1986},
\begin{align}
	\mathbf{Q}\cdot \vec{t} &= \vec{f} , \nonumber\\
	\mathbf{C}\cdot\vec{u} &= \vec{e},
\end{align}
where $\vec{t},\vec{f},\vec{u}, \vec{e}$ are vectors denoting tension on struts, total force on sites, site displacements, and struts extensions, respectively.  For a frame containing $N$ hinges and $E$ struts, vectors $\vec{t}$ and $\vec{e}$ are $E$ dimensional, and vectors $\vec{f}$ and $\vec{u}$ are $Nd$ dimensional where $d$ is the dimension of space and we take $d=2$ for all of our discussions.

Vectors in the null space of $\mathbf{Q}$ represent equilibrium distributions of tensions on struts that result in no net force on hinges, called states of self stress (SSS).  Vectors in the null space of $\mathbf{C}$ represent hinge displacements that do not change the length of any strut, called zero modes.  
A subset of these zero modes, excluding trivial ones for rigid translations and rotations of the whole frame, are called FMs or mechanisms, which denote relative displacements of hinges (deformations).

It is straightforward that the two matrices $\mathbf{Q}$ and $\mathbf{C}$ are transpose of one another---in fact, both of them are simply determined by the directions of the struts.  Therefore they must have the same rank.  Applying rank-nullity theorem to these matrices leads to the Maxwell-Calladine index theorem
\begin{align}
	\nu\equiv N_0 - N_s = 2N-E
\end{align}
where $\nu$ is the Maxwell-Calladine index, $N_0$ and $N_s$ are the numbers of zero modes and SSSs in the frame.  An intuitive way to understand this equation is that, when an additional strut is introduced to the frame ($E$ increases by 1), it is either a new constraint and eliminates one zero mode, or it is a redundant constraint, and introduces a new SSS. 

For a frame to be able to support certain external stress (e.g. shear or compression), the stress must overlap with at least one SSS, because SSSs describe the complete linear space of possible distribution of stresses in the frame leaving all nodes in force balance.  For this reason, the relation between SSSs and reciprocal diagrams has been exploited in the literature to study interesting problems such as jamming of granular particles~\cite{Tighe2010,DeGiuli2011,sarkar2013origin,Thomas2018} or cell sheets with active tensions~\cite{noll2017active}.

The number of FMs in a finite frame with open boundary conditions (no external forces on hinges) is then given by
\begin{align}\label{EQ:Nmfinite}
	N_m =2 N-E+N_s - 3
\end{align}
where we subtract the three trivial zero modes of rigid translations and rotation.  

This formulation [except Eq.~\eqref{EQ:Nmfinite}] applies equally to periodic structures, where the relation between number of zero modes $n_0 (\vec{q})$ and SSSs $n_s (\vec{q}) $ at each momentum point $\vec{q}$ is given by
\begin{align}
	\nu(\vec{q}) =  n_0 (\vec{q}) - n_s (\vec{q}) = 2n-e ,
\end{align}
where $\nu(\vec{q})$ is the Maxwell-Calladine index at momentum $\vec{q}$ and  
$n,e$ are the number of hinges and struts in each unit cell.  Under periodic boundary conditions, there is no trivial rotational zero mode, and there are 2 trivial translational zero modes only live at $\vec{q}=0$.

With this formulation we can now consider the existence of reciprocal diagrams.  
A finite frame has a reciprocal diagram only when it has a SSS,
\begin{align}
	N_s >0.
\end{align}
If $N_s>1$ the frame has multiple reciprocal diagrams and they form a linear space as we discuss more below.

For periodic lattices, SSSs at different $\vec{q}$ can be used to generate reciprocal diagrams at different unit-cell sizes.  In particular, Maxwell lattices (i.e., $\langle z \rangle =4$ so $2n=e$) have 
\begin{align}\label{EQ:NsLatt}
	n_s (\vec{q}=0) \ge 2
\end{align}
from the 2 trivial translational zero modes at $\vec{q}=0$, so a Maxwell lattice must have at least 2 reciprocal diagrams with the periodicity same as its own.  An example illustrating this is shown in Fig.~\ref{FIG:reciprocalOne}(c).

Further interesting results can be derived concerning the number of reciprocal diagrams.  
For a finite frame, the number of hinges ($N$), struts ($E$), and faces ($F$) are related by Euler's formula
\begin{align}
	N + F - E =2 .
\end{align}
The reciprocal diagram has
\begin{align}
	N^* = F, \quad F^*=N, \quad E^*= E,
\end{align}
and these numbers are also related by the Euler's formula.  One then has a relation between the numbers of FMs and SSSs of finite reciprocal diagrams~\cite{mitchell2016mechanisms}
\begin{align}\label{EQ:IndexFinite}
	(N_m - N_s) + (N_m^* - N_s^*) =-2 .
\end{align}
The perhaps simplest case of a pair of reciprocal diagrams is then $N_s = N_s^* =1$ and $N_m=N_m^*=0$.  The reciprocal diagrams pair shown in Fig.~\ref{FIG:reciprocalOne}(a) is such an example.  

Another interesting case is when a diagram $A$ has $N_s=2$ and $N_m=0$.  The two SSSs can be used to generate two different reciprocal diagrams $A^*\perp A$ and $A^*_2 \perp A$.  Because both $A^*$ and $A^*_2$ are reciprocal to $A$, there is a simple mapping between $A^*$ and $A^*_2$ such that face $\leftrightarrow$ face, node $\leftrightarrow$ node, and edge $\leftrightarrow$ edge, and the corresponding edges in $A^*$ and $A^*_2$ are parallel to one another.  We denote this relation as $A^* \parallel A^*_2$.  According to Eq.~\eqref{EQ:IndexFinite} we have $ N_m^*-N_s^*=0$ for both $A^*$ and $A^*_2$.  Naively one might expect $A^*$ and $A^*_2$ to be ``isostatic'' (i.e. $N_m^*=N_s^*=0$), but this can not be true because each of them  has at least one SSS that leads to $A$.  
We show such an example in Fig.~\ref{FIG:reciprocalOne}(b) where $N_m^*=N_s^*=1$ for $A^*$ and $A^*_2$.  There is actually a deep relation between these diagrams with multiple SSSs are present, and we discuss more on this below in Sec.~\ref{SEC:DualityTheorem}.

It is worth noting that counting the number of faces in a diagram is not always trivial.  If the diagram is planar graph (which is the case for most soft matter problems), the count is obvious and one just needs to add the exterior face for finite diagrams, which is already included in Euler's formula.  This also works for nonplanar diagrams that are projections of a spherical polyhedron [such as $A$ in Fig.~\ref{FIG:reciprocalOne}(b)], and the face counting is according to the polyhedron.  A mathematically rigorous discussion of this reciprocal relation can be found in Refs.~\cite{crapo1993plane,crapo1994spaces}.

For periodic lattices, a similar derivation leads to
\begin{align}
	 \lbrack n_0(\vec{q}) - n_s(\vec{q}) \rbrack + \lbrack n_0^*(\vec{q}) - n_s^*(\vec{q})\rbrack =0 .
\end{align}
for every $\vec{q}$.  One can also write this relation as
\begin{align}\label{EQ:IndexLatt}
	\nu(\vec{q})  = -\nu^* (\vec{q}) .
\end{align}
Note that the original lattice and its reciprocal may have different primitive vectors, and the momentum $\vec{q}$ here is measured in unit of the lattice's own reciprocal vectors.  In other words, when using this equation one has to bear in mind that $\vec{q}$ denote wave numbers rather than actual lengths in momentum space.  In addition, it is the number of zero modes $n_0(\vec{q})$ that enters this formula, rather than the number of FMs which is the case for the finite frames in Eq.~\eqref{EQ:IndexFinite}.

Thus, \emph{reciprocal periodic lattices have opposite Maxwell-Calladine indices at every momentum, and thus the reciprocal diagram of a Maxwell lattice must also be a Maxwell lattice.  }

Equations~(\ref{EQ:IndexFinite},\ref{EQ:IndexLatt}) tell us that the numbers of zero modes and SSSs between reciprocal diagrams relate in an opposite way.  We show below that, in fact, their relation is beyond this: there is a geometric mapping between each SSS in a frame and each FM in its reciprocal diagram, and vice versa. 

\subsection{Mechanical duality between reciprocal diagrams}\label{SEC:DualityTheorem}
We are now ready to introduce the mechanical duality theorem that is central for our discussion of topological mechanics in parallelogram tilings:

\textbf{Theorem:} \emph{For any pair of reciprocal diagrams, there is a one-to-one mapping between each state of self stress in one diagram (excluding the one that generates the reciprocal diagram under consideration) and each floppy mode of the reciprocal diagram.}

This theorem has appeared in the literature in different forms~\cite{crapo1994spaces,mitchell2016mechanisms}.  Below we give our version of the proof, which we believe is a somewhat easier-to-read version for the condensed matter community.  For a more mathematically rigorous statement and proof of this theorem (e.g., on degenerate diagrams) please see Ref.~\cite{crapo1994spaces}.

To prove this theorem, we start by considering a diagram $A$ which has two linearly independent SSSs [see, e.g., Fig.~\ref{FIG:reciprocalOne}(b)].  As we discussed above in Sec.~\ref{SEC:CountDual}, they lead to two different diagrams $A^* \parallel A^*_2$ which are both reciprocal to $A$.  Because corresponding edges in $A^*$ and $A^*_2$ are parallel to one another, the node displacements $\vec{v}_{\alpha^*}$ that lead from $A^*$ to $A^*_2$ must satisfy
\begin{align}
	(\vec{v}_{\alpha^*}-\vec{v}_{\beta^*}) \times \hat{b}_{i^*} =0 ,
\end{align}
where $\hat{b}_{i^*}$ is the unit vector pointing along edge $i^*$ which connects nodes $\alpha^*,\beta^*$ in diagram $A^*$. In other words, these displacements must be ``irrotational''.  

Next we show that, having $A^* \parallel A^*_2$ is equivalent to the fact that diagram $A^*$ has a FM.  We define a new set of node displacements $\vec{u}_{\alpha^*}$ which are $\vec{v}_{\alpha^*}$ rotated by $\pi/2$ at each node $\alpha^*$.  We then have
\begin{align}
	(\vec{u}_{\alpha^*}-\vec{u}_{\beta^*}) \cdot \hat{b}_{i^*} =0 ,
\end{align}
so $\vec{u}_{\alpha^*}$ do not change the length of any edge, and is a FM of $A^*$.  The converse is also true that from any FM of a diagram one can build a parallel diagram.  
Figure.~\ref{FIG:reciprocalOne}(c) shows an example of applying this theorem to periodic Maxwell lattices.

Therefore the mechanical duality theorem is proven, because from any additional SSS of a diagram $A$ one can construct an additional reciprocal diagram $A^*_2$ which is parallel to the first reciprocal diagram $A^*$, and yields a FM of $A^*$.  On the other hand, if a diagram has a FM, it has a parallel diagram, which is also reciprocal to its reciprocal diagram, yielding an additional SSS of the reciprocal.

This theorem indicates many interesting properties of frames, especially Maxwell networks.  For example, \emph{for two Maxwell periodic lattices that are reciprocal diagrams of one another [e.g., in Fig.~\ref{FIG:reciprocalOne}(c)], if one of them is topologically polarized, the other one must exhibit a topological polarization in the opposite direction}.  The reason is that starting from a boundary FM of lattice $A$, one can construct $A_2' \parallel A$ [the prime denotes that it's not a homogeneous lattice as $A_2$ in Fig.~\ref{FIG:reciprocalOne}(c)], and thus a boundary SSS of the reciprocal lattice $A^*$ (from the difference between $A$ and $A_2'$).  Thus, the FMs of  $A^*$ must be exponentially localized at the opposite boundary because solutions to $\det \mathbf{Q}(\vec{q})=0$ and $\det \mathbf{C}(\vec{q})=0$ have opposite imaginary parts of momentum.  Therefore, $A$ and $A^*$ have opposite topological polarizations.  The topological mechanics of Penrose tiling discussion we have below is a manifestion of this relation in quasicrystalline structures.

\section{Boundary floppy  modes in modified parallelogram-tilings}

\subsection{Mechanical duality between parallelogram tilings and fiber networks}
Floppy modes in original parallelogram-tilings are all bulk modes, as we discussed in Sec.~\ref{SEC:BulkModes}.  In this section, we discuss how boundary FMs can be introduced in parallelogram-tilings through infinitesimal geometric changes, exploiting the dual mechanics between parallelogram-tilings and fiber networks.

\begin{figure*}[t]
\centering
\includegraphics[width=1\textwidth]{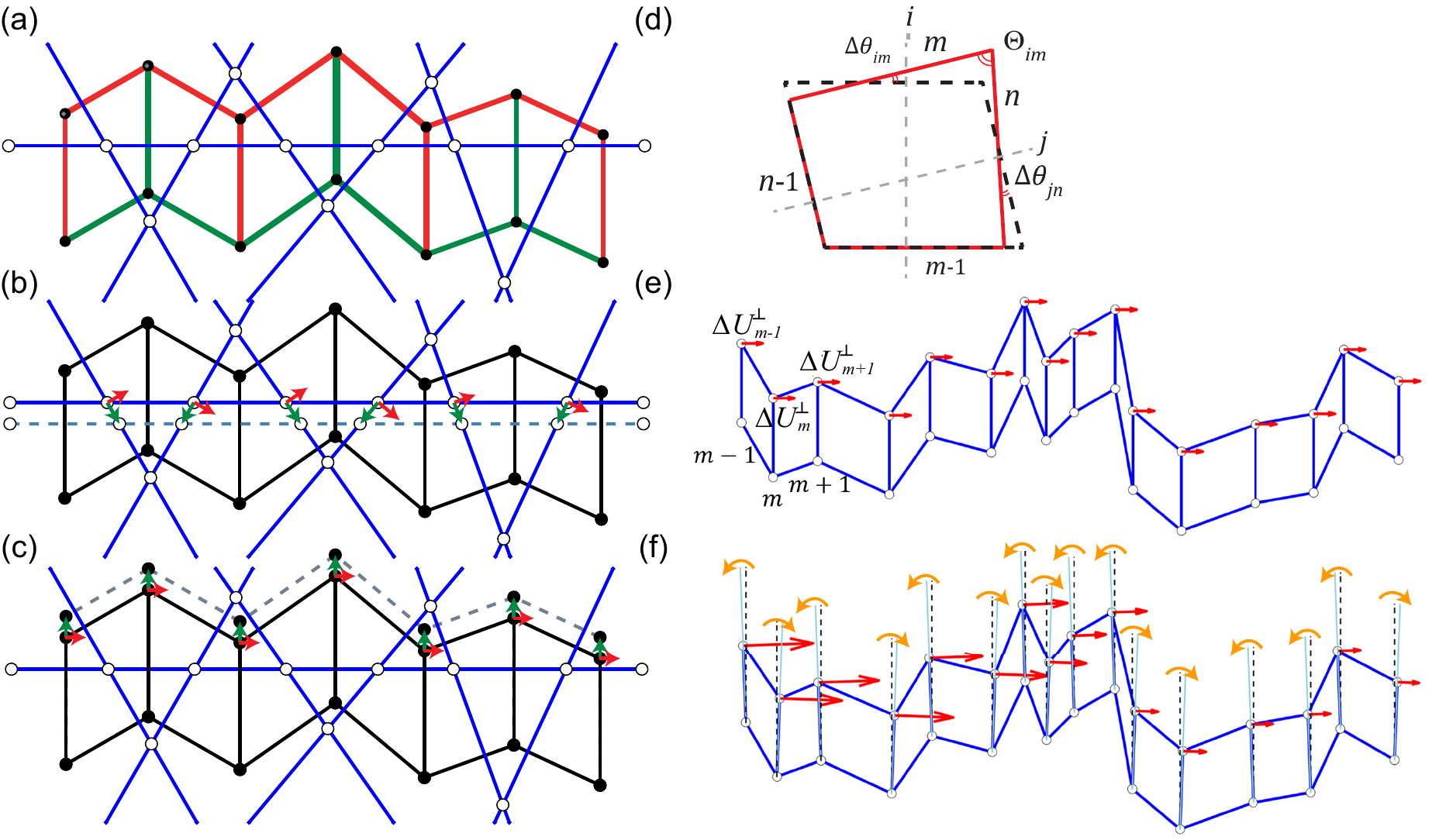}
\caption{The reciprocity relation between parallelogram tilings and fiber networks.  (a) A strip in a parallelogram tiling carries a SSS (sign of tension shown by red and green colors and magnitude shown as edge thickness).  Linear combinations of these SSSs from different strips lead to its reciprocal fiber network (blue lines with white crosslinking nodes).  (b) A different linear combination of SSSs of the parallelogram tiling leads to a different fiber network, which is parallel to other fiber networks that are reciprocal to the same parallelogram tiling. (The definition of parallel relation between diagrams is introduced in Sec.~\ref{SEC:CountDual}.)  Green and red arrows show irrotational and FM displacements $\vec{v}$ and $\vec{u}$ relating these fiber networks.  
(c) Similarly, different linear combinations of SSSs of a fiber network leads to different parallelogram tilings that are parallel to one another.  (d) Illustration for the dual transfer matrix at a quadrilateral (red solid lines).  Gray dashed lines show the corresponding fibers of the strips $i$ and $j$ that cross at this quadrilateral.  Black dashed lines denote the corresponding parallelogram where $\Delta\theta_{i,m}=\Delta\theta_{j,n}=0$.
(e)  A strip $i$ of parallelograms carries a bulk FM where $\Delta U_{i,m}^{\perp}$ (red arrows) is the same at every vertical edge.  (f) After small rotations of every vertical edge in directions such that $\Delta\theta_{i,m}\cot\Theta_{i,m}<0$ at each modified parallelogram (black dashed lines show the vertical direction,  light blue solid lines show extended direction of the rotated edges so the rotation is more visible, and orange arrows show the direction of rotation), the FM decays from left to right, as determined in Eq.~\eqref{EQ:FMsmall}.
}\label{FIG:FiberTilingDual}
\end{figure*}

It is straightforward to construct the Maxwell reciprocal diagram for an arbitrary parallelogram-tiling.  All faces in a parallelogram-tiling have four edges, corresponding to $z=4$ nodes in the reciprocal.  In addition, following a strip of parallelograms with parallel edges (same as the strip for the bulk FM discussion in Fig.~\ref{FIG:BulkMode}), we have a straight line perpendicular to these edges in the reciprocal.  Therefore it is clear that the reciprocal of any parallelogram-tiling leads to a fiber network with straight fibers and two fibers crossing at each node, as shown in Fig.~\ref{FIG:FiberTilingDual}(a).  Note that this reciprocal relation is based on an infinitely large parallelogram-tiling where the corresponding fibers do not terminate.  For a finite tiling or a finite fiber network, proper boundary forces have to be added for the reciprocal relation to hold, or one can define ``quasi'' reciprocal relations as we discuss in Sec.~\ref{SEC:TopoQC}.

The existence of this reciprocal relation already carries interesting information: all parallelogram-tilings must have SSSs in order to have their fiber network reciprocals, and all fiber network must have SSSs in order to have their tiling reciprocals, in the infinite size case.  In fact, each strip in a parallelogram-tilings carries a SSS [Fig.~\ref{FIG:FiberTilingDual}(a)], and each fiber in the fiber network carries an SSS where every segment carries the same tension.  Because the numbers of SSSs in each parallelogram tiling and each fiber network are subextensive (equal to the number of strips and fibers respectively), one can make linear combinations of these SSSs to generate multiple reciprocals, and these reciprocals are related to one another by (bulk) FMs, as shown in Fig.~\ref{FIG:FiberTilingDual}(b-c).  This follows directly from the mechanical duality theorem we discussed in Sec.~\ref{SEC:DualMech}, and their numbers are related by
\begin{align}\label{EQ:IndexInfi}
	\nu+\nu^* \equiv (N_0 - N_s) + (N_0^* - N_s^*) =0 ,
\end{align}
where we consider  infinite tilings and fiber networks on tori.  Note that $N_0$ is the number of zero modes instead of the FMs in this equation.  
In fact,
\begin{align}
	N_0 = N_s = N_0^* =  N_s^* = N_F ,
\end{align}
where $N_F$ is the number of fibers in the fiber network (which wraps around the torus), because each fiber and each strip carries a SSS and a zero mode, as we show in Fig.~\ref{FIG:FiberTilingDual}(a-c).
  
\subsection{Boundary floppy modes in fiber networks}
From the mechanical duality between parallelogram-tilings and fiber networks we discussed above, it is not difficult to realize that if we perturb the geometry of the fiber network a little bit, for example, following the construction in Ref.~\cite{Zhou2018}, to polarize its FMs and SSSs, the corresponding SSSs and FMs in the reciprocal parallelogram-tiling (after corresponding geometric perturbations) will also polarize and become boundary modes, albeit on opposite boundaries as in the fiber network.

Here we first briefly review the construction in Ref.~\cite{Zhou2018} to polarize fiber networks and introduce boundary FMs.  Using a transfer matrix method which exactly calculate the FM displacements along nodes on a fiber, it was found that the longitudinal projections of the FM displacements obey the following equation
\begin{align}\label{EQ:FiberTM}
	U_m = \lbrack 1-\Delta \theta_m \cot \Theta_m + O(\Delta \theta_m^2)\rbrack U_{m-1}
\end{align}
where $m$ labels the nodes along the fiber under consideration (the direction of which we define to be left to right as $m$ increases, without losing generality), $U_m$ is the FM displacement on node $m$ projected along the fiber segment right to node $m$, $\Delta \theta_m$ is the bending angle of the fiber at node $m$, and $\Theta_m$ is the angle between the fiber under consideration and the fiber that crosses this one at node $m$.  It is easy to see that if $\Delta \theta_m=0$, corresponding to the fiber being straight at node $m$, the FM projection keeps the same projection from node $m-1$ to $m$.  If the fiber is straight everywhere, it carries a bulk FM without decay, as we show in Fig.~\ref{FIG:FiberTilingDual}(b).  

When the fiber bends at the nodes, the FM displacement projection is no longer a constant along the fiber, instead they evolve according to Eq.~\eqref{EQ:FiberTM}, which is a leading order equation at small bending angles $\Delta \theta_m$.  If the bending of the fiber is such that $\langle U_m / U_{m-1}\rangle <1$ (where $\langle \cdots \rangle$ represent disorder average), the FM decays from left to right on the fiber, and the SSS grows from left to right.  The transfer matrix for the SSS on the fiber was not directly discussed in Ref.~\cite{Zhou2018}.  Below we introduce a transfer matrix for FMs in parallelogram-tilings, which also describes the localization of SSS along a fiber, according to the mechanical duality theorem.

\subsection{Constructing boundary floppy  modes in parallelogram-tilings}
Bending fibers in a fiber network renders the FM and SSS on that fiber to be localized on opposite tips of the fiber, as we reviewed above.  Following the mechanical duality theorem, we could find the  modified parallelogram-tiling (where ``modified'' refers to the fact that some parallelograms in the tiling are changed into general quadrilaterals where edges are not parallel) that is reciprocal to  the modified fiber network (where ``modified'' refers to the fact that some fibers are bent).  Because of the mechanical duality theorem, from the polarization of FMs and SSSs in the modified fiber network, the corresponding parallelogram-tiling must also have polarized SSSs and FMs.

Here we take a somewhat different route, by directly introducing a ``dual transfer matrix'' for FMs in modified parallelogram tilings, because it is not computationally straightforward to find the reciprocal of a modified fiber network.  The dual transfer matrix describes the evolution of the FM rotational component of displacements along a strip of modified parallelogram (which are dual to edge tensions in the reciprocal fiber network, as we discussed in Sec.~\ref{SEC:DualityTheorem}).  For each quadrilateral in the tiling, the FM displacements must satisfy the following relation because the deformed quadrilateral is still a closed polygon [see Fig.~\ref{FIG:FiberTilingDual}(d)]
\begin{align}\label{EQ:Quadrilateral}
	\Delta \vec U_{i,m-1}+\Delta \vec U_{j,n-1}=\Delta \vec U_{i,m}+\Delta \vec U_{j,n} 
\end{align}
where $i,j$ label the two strips that cross at the quadrilateral, $m,n$ labels edges along strips $i,j$ respectively.  $\Delta \vec U_{i,m}$ is the difference between the displacement vectors of the two nodes connected by edge $(i,m)$.  Because FMs do not extend edges, these vectors $\Delta \vec{U}$ only have components perpendicular to the edge, 
\begin{align}
	\Delta U_{i,m}^{\perp} \equiv \Delta \vec{U} \cdot \hat{e}_{i,m}^{\perp}
\end{align}
where $\hat{e}_{i,m}^{\perp}$ is the unit vector perpendicular to edge $(i,m)$ which is along direction $\theta_{i,m}$.  These perpendicular components of Eq.~\eqref{EQ:Quadrilateral} leads to two equations (because $\hat{e}^{\perp}$ is different for each edge around the quadrilateral, it is still a vector equation with two components), which we use to solve for $(\Delta U_{i,m}^{\perp},\Delta U_{j,n}^{\perp})$ as a function of $(\Delta U_{i,m-1}^{\perp},\Delta U_{j,n-1}^{\perp})$, yielding the dual transfer matrix for FMs in modified parallelogram tilings,
\begin{eqnarray}\label{C3}
\left(
\begin{array}{c}
\Delta U_{i,m}^{\perp}\\
\Delta U_{j,n}^{\perp}
\end{array}
\right)= M_{i,m;j,n} \cdot
\left(
\begin{array}{c}
\Delta U_{i,m-1}^{\perp}\\
\Delta U_{j,n-1}^{\perp}
\end{array}
\right)
\end{eqnarray}
where the transfer matrix
\begin{eqnarray}\label{10}
M_{i,m;j,n} = \left(\begin{array}{cc}
\frac{\sin(\Theta_{i,m}+\Delta\theta_{i,m})}{\sin\Theta_{i,m}} & \frac{\sin\Delta\theta_{j,n}}{\sin\Theta_{i,m}}\\
-\frac{\sin\Delta\theta_{i,m}}{\sin\Theta_{i,m}} & \frac{\sin(\Theta_{i,m}-\Delta\theta_{j,n})}{\sin\Theta_{i,m}}
\end{array}\right)
\end{eqnarray}
where $\Delta\theta_{i,m} = \theta_{i,m}-\theta_{i,m-1}$ and $\Delta\theta_{j,n}=\theta_{j,n}-\theta_{j,n-1}$ are the angles that describe how much the quadrilateral deviate from a parallelogram (corresponding to ``bending angles'' in the reciprocal fiber network), and $\theta_{j,n}-\theta_{i,m}=\Theta_{i,m}$ is the angle between edge $(i,m)$ and the neighboring edge $(j,n)$ (corresponding to the ``intersecting angle" in the reciprocal fiber network).

In the special case where all $\Delta \theta =0$, corresponding to all perfect parallelograms in the tiling, $\Delta U_{i,m}^{\perp}$ and $\Delta U_{j,n}^{\perp}$, which we call ``FM edge rotations'' in the following discussion, simply transmit along the two strips $i$ and $j$ with no mixing, giving rise to bulk modes as shown in Fig.~\ref{FIG:BulkMode}.  This is equivalent to a fiber network with all straight fibers.  

When the parallelogram edges are rotated and deviate from this state the FMs also changes, similar to a fiber network with bent fibers.   To study the growth and decay of the FM displacements and find out the geometry that localizes FMs, we further make an approximation that the angle changes $\Delta \theta$ are small, meaning that the modified parallelogram-tiling is not too different from the original one.  In this limit, it is straightforward to see that the two directions decouple and we have the equation for the FM edge rotations along strip $i$
\begin{eqnarray}\label{EQ:FMsmall}
 \Delta U_{i,m}^{\perp} = [1+\Delta\theta_{i,m}\cot\Theta_{i,m}+\mathcal{O}(\Delta\theta^2)]\Delta U_{i,m-1}^{\perp}\qquad
\end{eqnarray}
and an equation of the same form applies to strip $j$.  Note that we have assumed that $\cot\Theta$ does not diverge in this expansion, which is satisfied in tilings where the parallelogram angles are not too close to $0$ (naturally satisfied in Penrose tiling).  
From this, it is clear that if the edges are rotated in a coherent way such that $\Delta\theta_{i,m}\cot\Theta_{i,m} \ge 0$ ($\le 0$) at most parallelograms along the strip, the FM edge rotation $\Delta U^{\perp}$ coherently grows (or decays) along the strip.  We show such an example of an isolated strip in Fig.~\ref{FIG:FiberTilingDual}(e-f).  

It is worth pointing out that the same transfer matrix applies to SSSs on fiber networks, because of the mechanical dual theorem discussed in Sec.~\ref{SEC:DualityTheorem}.

In principle, one could use this transfer matrix to calculate FMs for any modified parallelogram-tilings.  The math is more involved than calculating FMs in modified fiber networks, because here it is the FM edge rotations $\Delta U^{\perp}$ that enter the transfer matrix, and to obtain the FM one needs to solve for the node displacement vectors.  It is a well defined problem given proper boundary conditions (similar to boundary conditions discussions in Ref.~\cite{Zhou2018}).

\subsection{Topological winding number of a strip in modified parallelogram-tilings}
A topological winding number can be defined for the localized FM on a strip, following a similar discussion of topological winding number for boundary FMs on bent fibers in Ref.~\cite{Zhou2018}.  To do this we need to first define a compatibility matrix, the null space of which describes the FM on a strip.  Because of the nature of the strip FM, as described by the dual transfer matrix we defined above, the compatibility matrix maps the edge rotations $\Delta U^{\perp}$ along strip $i$, instead of the node displacements, to edge extensions
\begin{eqnarray}\label{13}
\delta l_{i,b} = \sum_{m=1}^{N_i}C_{b m}\Delta U_{i,m}^\perp
\end{eqnarray}
where $N_i$ is the number of parallel edges in strip $i$. The compatibility matrix is given by 
\begin{eqnarray}\label{14}
C_{b m} =  \sin\Theta_{i,m} [\delta_{b,m}-(1+\Delta\theta_{i,m}\cot\Theta_{i,m})\delta_{b-1, m}] .\qquad
\end{eqnarray}
Here the edge extension $\delta l_{i,b}$ denotes the extension of the top edges of this strip [see Fig.~\ref{FIG:FiberTilingDual}(e-f)].  The set up is that the bottom of the strip is fixed and nodes on the top can move.  The component of the site displacements that are perpendicular to the vertical edges gives the edge rotations $\Delta U^{\perp}$, and the compatibility matrix describes how they extend the edges on the top.  One can also do this by fixing the top boundary and calculate bottom edge extensions, and the result will be the same.  In addition, it is easy to see that if a set of edge rotations satisfy the dual transfer matrix for FMs [Eq.~\eqref{EQ:FMsmall}], one gets $\delta l_{i,b} =0$ for all edges as expected.

We can then define a topological winding number for this strip using the momentum space form of this compatibility matrix $C(q_1,q_2)$ where the two momenta $q_1,q_2$ corresponding to real space labels $b$ for (top row) edge extensions and label $i$ for (vertical) edge rotations.  Note that this Fourier series is based on labeling quadrilaterals along the strip which have different sizes, and thus not homogeneous in space.  The winding number of strip $i$ is then defined as 
\begin{eqnarray}\label{15}
\mathfrak{N}_i = \frac{1}{N_i}\frac{1}{2\pi}\oint_0^{2\pi}dk\frac{d}{dk}{\rm Im}\,\ln\, \det C(q_1+k, q_2+k), \qquad\quad
\end{eqnarray}
which can take two values $0$, corresponding to FM localizes on the right, and $1$, corresponding to FM localizes on the left.  This winding number is only well defined when the strip does not have a bulk FM, so the phonon spectrum is gapped, and this corresponds to the case where not all quadrilaterals are parallelograms.  

The form of this winding number is the same as the one appeared in Ref.~\cite{Zhou2018} for disordered fiber networks.  A detailed discussion of how this winding number controls the localization of a FM in a 1D disordered chain, and why it only takes values 0 and 1, can be found in Ref.~\cite{Zhou2018}.  This winding number is an extension of the winding number defined in Ref.~\cite{Kane2014} to non periodic systems.

\section{Topological mechanics of the Penrose tiling}\label{SEC:TopoQC}
In previous sections we have shown that in an arbitrary tiling of parallelograms, rotating edges by small angles can localize FMs and SSSs on opposite ends of strips of parallelograms, and we name these tilings after the geometric changes modified parallelogram tilings.  The exponential localization of these FMs and SSSs, when the parallelogram shape changes are coherent along the strips, is topological and described by a winding number as defined in Eq.~\eqref{15}.

In this section, we discuss how a 2D ``topological polarization vector'' $\vec{R}_T$ can be defined in quasicrystalline parallelogram tilings, and we use the Penrose tiling as an example.

First, there are five families of strips in a Penrose tiling, perpendicular to the five star-vectors $\{ \vec{a}_i \}$ that were used in generating the tiling in the GDM.  We define these perpendicular directions to be $\{ \vec{a}_i^{\perp} \}$ which are along $\pi/2, 9\pi/10, 13\pi/10, 17\pi/10, \pi/10$.

Naively, one may try to pick one family of strips and polarize them, leaving all other families of strips unpolarized, to obtain a state with $\vec{R}_T$ pointing along these strips ($\vec{a}_i^{\perp}$).  However, the operation that polarizes a strip can not be isolated: it involves rotating edges perpendicular to the strip.  When edges in a strip are rotated, nodes are displaced to arrive at the modified parallelogram-tiling, in which edges in other strips are necessarily rotated as well.

Thus, we choose to pick a subset of nodes in the Penrose tiling and give them small displacements, and study the topological polarization of the resulting modified Penrose tilings.  When a node is displaced, all edges connecting to it are rotated, which affects the FMs associated with the strips that pass through these edges.  

For example, we focus on the $z=5,6,7$ vertices shown in Fig.~\ref{FIG:TopoQCExample}(b), and consider displacing the center nodes of these vertices by a small amount $\vec{r}_0$.  [Note that there are two types of $z=5$ vertices in the Penrose tiling and we only displace the type shown in Fig.~\ref{FIG:TopoQCExample}(b).]  Using the FM transfer matrix equation~\eqref{EQ:FMsmall} we find how $\vec{r}_0$ polarizes all the strips.  In Fig.~\ref{FIG:TopoQCExample}(c) we show the polarization of each strip when $\vec{r}_0$ points to different ranges of angles.  These $z=5,6,7$ vertices appear in the Penrose tiling in different orientations, and the corresponding polarization phase appear to be the all the same for these vertices at different orientations.  By displacing all these vertices we obtain topologically polarized Penrose tilings shown in Fig.~\ref{FIG:TopoQCExample}.  One can also choose to perturb the geometry in other ways to polarize the tiling.  We show topological polarization phase diagrams for displacements of all vertices in the Penrose tiling in App.~\ref{APP:VERTEX}.

In a quasicrystalline tiling of parallelograms we do not have 2D unit cells.  Thus, upon taking care of the gauge choice of unit cells along each strip, we define the topological polarization of the modified Penrose tiling (which we call ``topologically polarized Penrose tilings'' now)
\begin{align}\label{EQ:RT}
	\vec{R}_T = \sum_{i=1}^{5} n_i \vec{a}_i^{\perp}
\end{align}
where $n_i=1$ if the FM is localized at the tip of the strip where $\vec{a}_i^{\perp}$ points to, and $n_i=-1$ if the FM is localized at the opposite tip of the strip.  

If $n_i$ are allowed to take independent values at each strip, there are $2^5=32$ different values of $\vec{R}_T$.  Our study based on displacing nodes to obtain topologically polarized Penrose tilings realized 10 of these possible choice, as shown in Fig.~\ref{FIG:TopoQCExample}(c).  These $\vec{R}_T$ directions, in stark contrast to symmetry directions in periodic lattices, originate from the special symmetry of the quasicrystalline Penrose tiling.

\begin{figure}[h]
\centering
\includegraphics[width=0.2\textwidth]{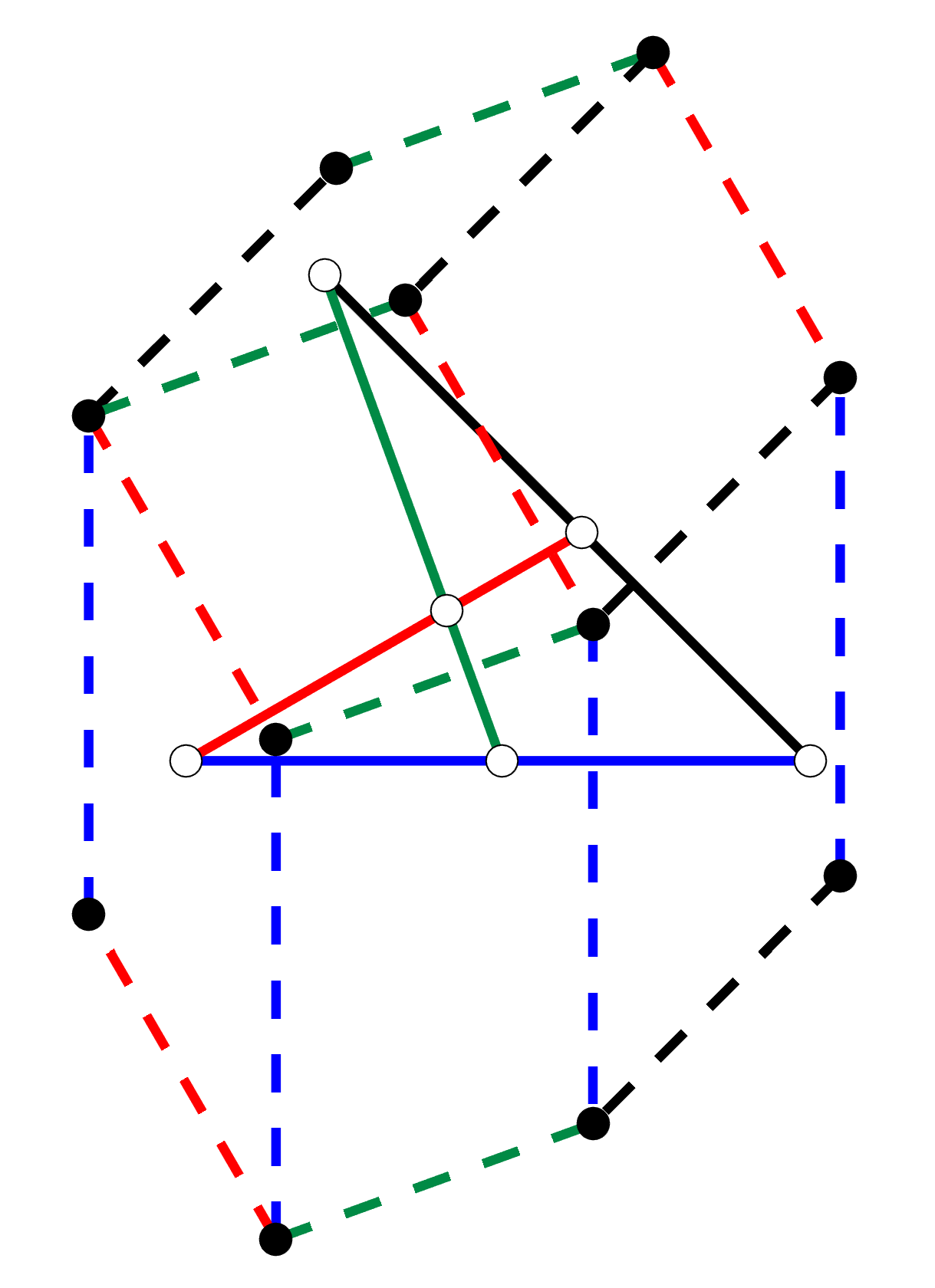}
\caption{The quasi-reciprocal relation between a finite parallelogram tiling and a finite fiber network ($N_F=4$).  Corresponding edges are shown in the same color (solid for the fiber network and dashed for the parallelogram tiling).  In this figure, $V_{\textrm{fiber}}=F_{\textrm{tiling}}=6, E_{\textrm{fiber}} =8, E_{\textrm{tiling}}=16$ so the number of FMs in the parallelogram tiling is $N_m=3=N_F-1$.}\label{FIG:QuasiDual}
\end{figure}

This topological polarization tells us the number of localized FMs on boundaries when we make a cut on the tiling.  To avoid trivial FMs generated by dangling ends we define a proper cut to be one such that all parallelograms share edge with at least two neighbors.  As a result the corresponding fiber network has no dangling ends, i.e., fibers only terminate at crosslinks.  It is worth noting that the corresponding finite parallelogram tilings and finite fiber networks are only ``quasi'' reciprocals of one another, because nodes on the boundaries of the parallelogram-tilings do not map to faces in the fiber networks, and thus one can not apply Eq.~\eqref{EQ:IndexFinite} to them to find the number of FMs.  A pair of such quasi-reciprocals are shown in Fig.~\ref{FIG:QuasiDual}.

Instead, the number of FMs is equal to the number of fibers in the quasi-reciprocal fiber network minus one.  We can see this from the following analysis.  Suppose there are $N_F$ fibers in the quasi-reciprocal fiber network and each fiber $i$ has $N_i$ nodes (crosslinks) on it. The total number of nodes in the fiber network is then
\begin{align}
	N_{\textrm{fiber}} = \frac{1}{2}\sum_{i=1}^{N_F} N_i
\end{align}
(note that each node lives on two fibers so we have the factor of $1/2$) and the total number of edges (fiber segments) is 
\begin{align}
	E_{\textrm{fiber}} = \sum_{i=1}^{N_F} (N_i-1) .
\end{align}
Thus the number of parallelograms in the quasi-reciprocal parallelogram tiling is (not including the exterior face)
\begin{align}
	F_{\textrm{tiling}} = \frac{1}{2} \sum_{i=1}^{N_F} N_i
\end{align}
and the number of edges is 
\begin{align}
	E_{\textrm{tiling}} = \sum_{i=1}^{N_F} (N_i+1) ,
\end{align}
note that each strip has two extra edges at the ends which have no corresponding edges in the fiber network.
Using Euler's theorem again (excluding the exterior face which is not a parallelogram in the tiling) we have the number of zero modes in the tiling
\begin{align}
	N_0 = 2N_{\textrm{tiling}} - E_{\textrm{tiling}} = N_F + 2 .
\end{align}
The number of FMs in the parallelogram tiling follows by removing the three trivial zero modes,
\begin{align}
	N_m = N_F-1.
\end{align}
It is straightforward from Sec.~\ref{SEC:BulkModes} that one FM is associated with one strip.  Out of all linear combinations of these strip bulk FMs, one is the global rotation.  For a large parallelogram tiling we can ignore this one and take $N_m \simeq N_F$.

Therefore, the number density of localized FMs on a cut boundary $s$ is
\begin{align}
	\tilde{\nu} = \rho_s \hat{n}_s \cdot (\vec{R}_T + \vec{R}_L^{s} )
\end{align}
where $\rho_s$ is the number density of terminating fibers on this boundary, $\hat{n}_s$ is the outward normal unit vector of the boundary, and $ \vec{R}_L^{s}$ is the dipole moment of the local count of FMs, as defined in Ref.~\cite{Kane2014}.  As a result, in Fig.~\ref{FIG:TopoQCExample}(a) the bottom left boundary perpendicular to $\vec{R}_T$ has no FMs and is as rigid as the bulk, whereas the top boundaries have more localized FMs than the number determined by $\vec{R}_L^{s}$.  To characterize this effect we calculate the weight of the zero modes on each node $\alpha$, defined as
\begin{align}
	\eta_{\alpha} \equiv \frac{1}{N_0} \sum_{s=1}^{N_0} \vert \vec{u}_{\alpha}^{(s)}\vert^2
\end{align}
where the sum is over all zero modes labeled by $s$, and $\vec{u}_{\alpha}^{(s)}$ is the displacement vector of mode $s$ on node $\alpha$.  The total number of zero modes $N_0$ normalizes this weight.  The sum includes the trivial 3 zero modes.  One could also choose to exclude them which merely results in $~O(1/N_0)$ change in the zero mode weight $\eta_{\alpha} $ on each node.

Exponential localization of FMs induces very asymmetric mechanical responses in opposite boundaries, an effect explored in Refs.~\cite{Rocklin2017,Zhou2018}.  Here we perform numerical simulations to measure local stiffness on opposite boundaries in a topologically polarized Penrose tiling with topological boundary FMs.  More details of our simulation can be found in App.~\ref{APP:Simulation}.  As shown in Fig.~\ref{FIG:TopoQCExample}(d-e), boundaries where $\vec{R}_T$ points toward show significantly lower local stiffness than the opposite edge, due to the exponentially localized topological FMs.

\section{Conclusion and Discussions}
In this paper we discuss topological mechanics in parallelogram-tilings and we are particularly interested in quasicrystalline tilings.  We show that with small geometric changes in node positions, FMs in parallelogram tilings can change from bulk modes to topological boundary modes.  Our construction works for both ordered and disordered tilings of parallelograms, and we discuss the particularly interesting class of quasicrystalline tilings and used the Penrose tiling to demonstrate our results.  

We find that the topological polarization $\vec{R}_T$ of the Penrose tiling is 10-fold symmetric, allowing rigid boundaries to show up in 10 different directions (Fig.~\ref{FIG:TopoQCExample}).  This is also true for quasicrystalline tilings with other symmetries, such as ones generated using 7-grids and 9-grids, that $\vec{R}_T$ can have 14-fold and 18-fold symmetries.  
This will open the door to designs of new quasicrystalline mechanical metamaterials where the topological modes have special symmetry properties beyond the crystallographic point group.

Our work extends topological mechanics to quasicrystalline structures.  The way we define the topological  polarization in quasicrystals is based on the topological winding number on each strip of parallelograms.  This winding number works the same way for quasiperiodic and disordered parallelogram strips.  Thus, the formulation of topological mechanics we discussed also applies to completely disordered parallelogram tilings, which are reciprocal to disordered fiber networks.  Our construction shares similarities with various formulations to define topological invariant in disordered systems~\cite{Niu1985,kitaev2006anyons,Prodan2010,bianco2011mapping,mitchell2018amorphous}.

Our results for the Penrose tiling highlight the uniqueness of topological mechanics in quasicrystals. Periodic crystals such as topological kagome and square lattices have well defined topological polarization $\vec{R}_T$ for the entire bulk, but the point symmetry of $\vec{R}_T$ must obey the crystallographic point group~\cite{Kane2014,Lubensky2015,Rocklin2016}. Disordered networks such as the Mikado model and the random parallelogram tilings only have well defined topological polarization $\vec{R}_T$ for each 1D strip, through averaging over disordered configurations~\cite{Zhou2018}. We show that quasicrystals such as Penrose tilings have well defined $\vec{R}_T$ for the entire bulk, while the point symmetry of $\vec{R}_T$ is the same as that of the quasicrystal, extending beyond the crystallographic point group. The method we use to define $\vec{R}_T$ was originally developed for disordered networks, but instead of averaging over disordered configurations, we average over only those allowed by the quasicrystal symmetry, a very small set of local configurations. The results are therefore analytic and not affected by disorder.

It will be interesting future work to consider other definitions of topological polarization in a quasicrystalline tiling of parallelograms, taking advantage of their quasiperiodic translational order: they can be written as a sum of periodic structures, relating to crystals at higher dimensions, and are not completely random.  
Some interesting explorations of defining topological index in quasicrystals in photonic, phononic, and electronic systems can be found in Refs.~\cite{Kraus2012,rechtsman2013photonic,Madsen2013,kraus2013four,
bandres2016topological,rosa2018boundary,Huang2018}.
Studying the analog of Weyl points in quasicrystalline tiling is another interesting future direction.   

The mechanical duality theorem we reviewed in this paper was known to the mathematical and engineering community, but remains largely undiscussed in the condensed matter community.  It relates SSSs and FMs in reciprocal structures, as well as the Airy stress function and 3D polyhedral surfaces that projects orthogonally into the 2D structure (``liftings'')~\cite{crapo1994spaces,mitchell2016mechanisms}.  Besides helping us understanding topological mechanics in quasicrystalline parallelogram tilings, we believe that the full potential of this duality theorem has yet to be explored in many soft matter problems, such as jamming of granular particles, gelation in dense suspensions, and motility of cell sheets.

\section*{Acknowledgements}
The authors acknowledge useful discussions with Bryan G.~Chen and D.~Zeb Rocklin.  This work was supported in part by the National Science Foundation under grant NSF-EFRI-1741618.


\appendix
\begin{widetext}

\section{Generating the Penrose tiling using the GDM}\label{APP:GDM}
In this section we discuss the numerical procedure, following the GDM, that we use to generate the Penrose tiling we use in this paper.  As we discussed in the main text, we start from a 5-grid and map each  polygon face to each node in the Penrose tiling.  Each line in the 5-grid is labeled by $m_i$ where $i=1,\cdots,5$ denotes the 5 directions.  Every node in the 5-grid is associated with the pair of lines that cross at this node $\{m_i, m_j\}$.  We find the space in the other three directions $\{m_l\}$ where this node $\{m_i, m_j\}$ sits in, through the following equation
\begin{eqnarray}\label{F1}
m_l(m_i,m_j) = &\Big \lfloor \tau_{l,ij}^{(1)}(m_i+\gamma_i)+\tau^{(2)}_{l,ij}(m_j+\gamma_j) 
-\gamma_l +1 \Big \rfloor 
\end{eqnarray}
with $\tau_{l,ij}^{(1)} = -\frac{\sin(j-l)\theta_p}{\sin(i-j)\theta_p}$ and $\tau_{l,ij}^{(2)}=\frac{\sin(i-l)\theta_p}{\sin(i-j)\theta_p}$,  $\theta_\textrm{p} = 2\pi/5$, and the $\gamma_i$'s are the shift of the lines relative to the origin.  The first two terms gives the projection of this node in the star vector $\vec{a}_l$ direction.  With the shift of $-\gamma_l +1$ the floor function finds the space labeled by $m_l$ (which is between lines $m_{l-1}$ and $m_l$).  With this equation we find four polygons surrounding this node, $\{m_i,m_j,\{m_l\}\}$, $\{m_i+1,m_j,\{m_l\}\}$, $\{m_i,m_j+1,\{m_l\}\}$, and $\{m_i+1,m_j+1,\{m_l\}\}$.  These give four neighboring nodes (which surround the parallelogram corresponding to node $\{m_i, m_j\}$ in the fiber network) in the Penrose tiling through equation~\eqref{EQ:rGDM}.

We then scan through pairs from the $C_5^2 =10$ choices from the set $\{m_1, m_2, m_3, m_4, m_5\}$ and take a large range of $m$ values in each direction.  This finds all the polygons in the range of the fiber network we generate, although each polygon is scanned multiple times.  
The resulting Penrose tiling is cut into finite domains for our calculations of boundary FMs.  In particular, we take $\gamma_i = 0.6$ for all $i$, which satisfies the condition of $\sum_{i=1}^5 \gamma_i =$ integer  condition for Penrose tiling with the matching rules, as discussed in Ref.~\cite{de1981algebraic}.

\section{Topological polarizations}\label{APP:VERTEX}
The Penrose tiling can also be polarized by displacing other vertices.  In Fig.~\ref{FIG:OtherVertices} we show a complete list of vertices and their corresponding topological polarization phase diagrams.
\begin{figure*}[h]
   \centering
   \includegraphics[width=1\textwidth]{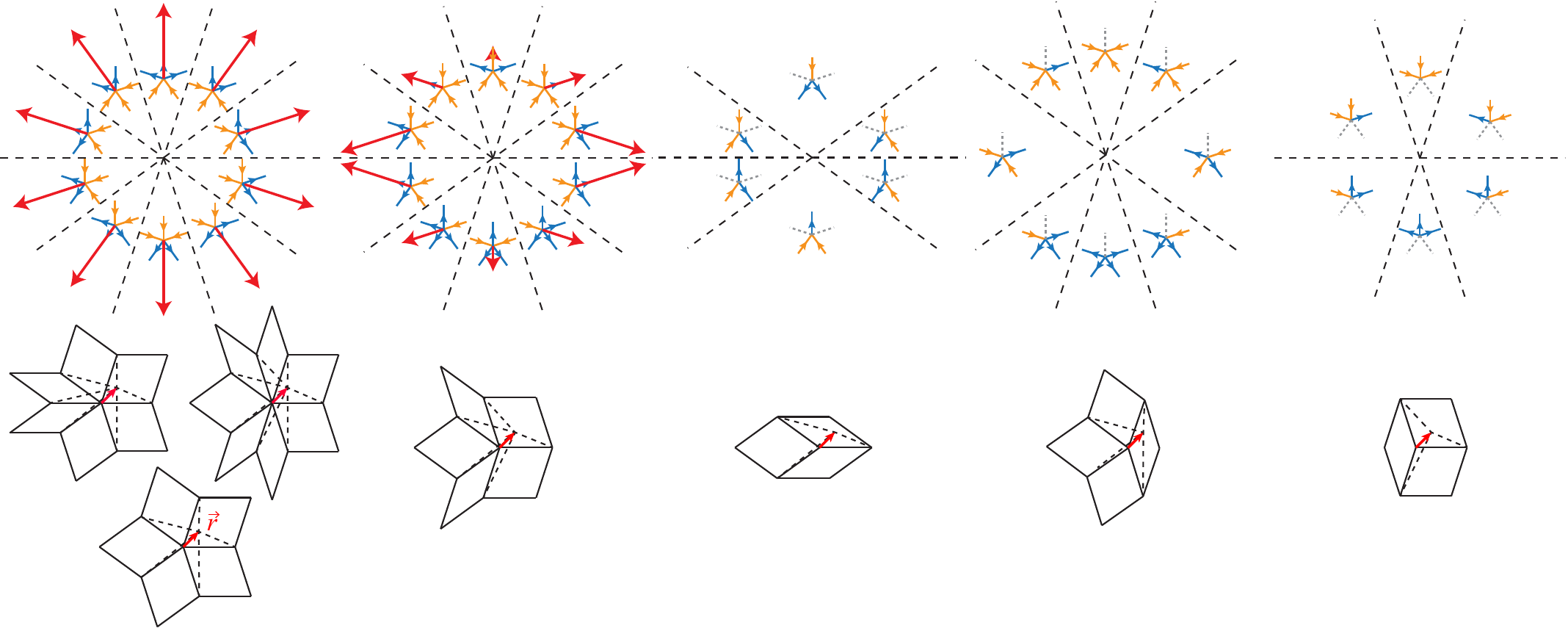} 
   \caption{A complete list of vertices in the Penrose tiling and their corresponding phase
diagrams of the topological polarization as a result of their displacements. We use the same convention here as in Fig.~\ref{FIG:TopoQCExample}. For $z = 3,4$ vertices, the FMs remain bulk modes on strips of parallelograms along some directions, because these strips do not pass through these $z = 3,4$ vertices. We denote these directions using gray dashed lines in the phase diagram.}\label{FIG:OtherVertices}
\end{figure*}

\section{Numerical measurement of local stiffness in modified Penrose tiling with topological boundary FMs}\label{APP:Simulation}
In the simulation, a finite-size Penrose tiling within a square box of $L = 40\times$length of edges, is prepared. We fix the left and right boundaries of the box and leave top and bottom free [Fig.~\ref{FIG:TopoQCExample}(d-e)].  
We apply in the direction $\hat{n}_s$ perpendicular to a boundary a small force $f\hat{n}_s$ on a site $i$ on the top or bottom boundaries. $\hat{n}_s = \hat{e}_y$ for the top and $\hat{n}_s = -\hat{e}_y$ for the bottom edge. 
The elastic energy is minimized by applying Molecular Dynamics with damping. We measure the displacement $\vec{u}_i$ of site $i$ due to $f$. The local stiffness at site $i$ is then calculated through $k_{\textrm{local},i} = f / (\vec{u}_i\cdot\hat{n}_s)$. We calculate the boundary local stiffness $k_{\textrm{local}}$ by averaging $k_{\textrm{local},i}$ over every site on the  boundary under consideration. The magnitude of $f$ is chosen to be small enough that the measurement is in the linear elasticity regime. We measure $k_{\textrm{local}}$ as a function of the vertex displacement $\vec{r}_0= r_0\hat{e}_y$ which defines the topologically polarized Penrose tiling, where the definition of $\vec{r}_0$ is shown in Fig.~\ref{FIG:TopoQCExample}(b). When $r_0>0$ ($r_0<0$), $\vec{R}_T$ is in $\hat{e}_y$ ($-\hat{e}_y$) direction and FMs are exponentially localized at the top (bottom) boundary.

\end{widetext}


%

\end{document}